\documentclass[11pt,aps]{revtex4-1}

\usepackage{graphicx}
\usepackage{wrapfig}
\usepackage{natbib}
\usepackage{xcolor}
\usepackage[normalem]{ulem}

\newcommand{\be}{\begin{equation}}
\newcommand{\ee}{\end{equation}}
\newcommand{\nn}{\mbox{} \nonumber \\ \mbox{} }
\newcommand{\ba}{\begin{eqnarray}}
\newcommand{\ea}{\end{eqnarray}}
\newcommand{\om}{\omega}
\newcommand{\Alfven}{ Alfv\'{e}n }

\newcommand\eg{{\it{{e.g., }}}}

\newcommand{\Lf}{{Lorentz factor}}

\newcommand{\Bf}{{magnetic field}}
\newcommand{\Bfs}{{magnetic fields}}
\newcommand{\Ef}{{electric  field}}
\newcommand{\Efs}{{electric fields}}

\newcommand{\NS}{neutron star}
\newcommand{\NSs}{{neutron stars}}
\newcommand{\EM}{electromagnetic}

\newcommand{\ms}{magnetosphere}
\newcommand{\mss}{magnetospheres}

\def\8{\infty}

\def\undertext#1{\vtop{\hbox{#1}\kern 1pt \hrule}}

\def\be{\begin{equation}}
\def\ee{\end{equation}}
\def\bea{\begin{eqnarray} & &}
\def\eea{\end{eqnarray}}

\begin{document}

\title{Fast Radio Bursts produced  during collapse of  macroscopic X-mode  in  magnetized pair plasma}
\author{Maxim Lyutikov \\
Department of Physics and Astronomy, Purdue University, \\ 525 Northwestern Avenue, West Lafayette, IN 47907-2036 }

\begin{abstract}
We demonstrate that in highly magnetized pair plasma nonlinear long-wavelength  X-modes experience wave collapse/breaking, whereby the wave undergoes severe spatial steepening, 
driven by nonlinear modifications of the    refractive index and strong ponderomotive forces. 
The collapse/wave breaking occurs in a narrow parameter regime, when the fluctuating part of the \Bf\ exceeds the guide field, and plasma magnetization is close to the current  starvation regime. This regime is naturally achieved in highly magnetized \NSs, magnetars. 
Breaking during a fraction of the dynamic timescale,  and quickly  generates high-k modes.  The initial EM energy, spread over large spatial scales,  is squeezed into these highly localized,  short-wavelength (yet macroscopic) singular pulses. The corresponding  \EM\ ``foam'' spectrum  is red, $E_k \propto k^{-2}$, while the particles' spectrum is exceptionally hard, $f(\gamma) \propto \gamma^0$  The wave collapse produces short bright EM pulses - astrophysical Fast Radio Bursts. The highest energy particles may produce short contemporaneous  high energy bursts. 
\end{abstract}

\maketitle

\section{Introduction}
Fast Radio
Bursts (FRBs) are millisecond-long bursts of radio emission coming from halfway across the Universe. At the peak, the (isotropic-equivalent) radio luminosity exceeds
billions of Solar luminosity  \citep{2007Sci...318..777L,2019A&ARv..27....4P,2019ARA&A..57..417C}. Understanding these phenomena, generation and propagation of ultra-intense \EM\ waves,  is a major problem in contemporary plasma/high energy  astrophysics. 

Detection of a radio burst from a Galactic magnetar by CHIME and STARE2 collaborations in coincidence with high energy bursts \citep{2020arXiv200510324T,2020arXiv200510828B,2020arXiv200506335M,2020arXiv200511178R,2020arXiv200512164T}, and the  similarity of its properties to the Fast Radio Bursts (FRBs),  give credence to the magnetar origin of FRBs.  The most compelling model, in our view, is the "Solar paradigm": generation of coherent radio emission during magnetospheric reconnection events \citep{2002ApJ...580L..65L,2013arXiv1307.4924P,2020arXiv200505093L,2023MNRAS.524.6024S,2022ApJ...934..140B}. 

In this work, we describe a  discovery of a highly promising mechanism for generating radio waves  by large-scale nonlinear X-modes in a highly magnetized plasma. The process is generally related to wave-overturn/collapse, but we are not aware of any close \EM\ analogues.
The key point is that in pair plasma mild ponderomotive effects, including self-excited by parametric-type processes,  lead to huge charge-neutral  density fluctuations. The resulting modifications of the basic plasma state lead to a number of surprising effects, \citep[\eg  Anderson self-localization of light in under-dense pair plasma][]{2025arXiv250920594L}. 

In what follows we first describe a pure plasma physics problem, a decay of a nonlinear X-mode perturbation in pair plasma. Later, in \S \ref{astro} we discuss astrophysical applications.

\section{The set-up and the code}

Consider  a nonlinear X-wave, of the same frequency and amplitude, counter-propagating perpendicular to the \Bf, the direction of propagation is $x$, \Ef\ along $y$,  \Bf, both the guide and the fluctuating are along $z$. 

Without addressing the history of the wave-wave interaction, we start at the moment when the corresponding \Efs\ subtract to zero, while \Bfs\  add. 
(Alternatively, such a configuration can be produced by a Harris sheet configuration, when the plasma density suddenly drops to zero, \eg, due to radiative cooling.)

We then model the initial configuration as a wave packet of Gaussian form; the initial \Bf\ then can be written as  
\be
B_y = B_0 \left (1- \delta  \times \mathcal{G} (x/\lambda) \right)
\label{By0}
\ee
where $\mathcal{G}$ is the Gaussian function. Parameter $\delta $ is the relative amplitude of the perturbation with respect to the guide field $B_0$. The \Bf\ (\ref{By0}) is treated as initial condition, not as an externally imposed field. (A particular,  $\pm x$ symmetric,  profile is chosen for purely numerical reasons, as "periodic" boundary conditions eliminate many simulation boundary-related issues.)

Importantly, there is no initial balancing plasma current, or plasma pressure (this then would be a double Harris-like current sheet). We just assume that external forces "pluck" the \Bf. This is a good approximation for X-mode (as opposed to \Alfven mode) in highly magnetized plasma. 

In the linear regime, $\delta \ll 1$, such initial set-up  launches two counter-propagating X-waves. As the nonlinearity parameter $\delta $ increases, a qualitatively new effect appears  in a limited regime  near $\delta \approx  2$, when the reversed field is of the order of the initial guide field.
(Our main investigation concerns the case of $\delta=2$, so that at $x=0$ the absolute value of the \Bf\ matches $|x| \gg 1 $ limit. This is just a nice symmetric example, not a particularly  odd/special case.)

As the magnetic ``string'' is released, it will generate \EM\ perturbations -  currents, and generally charge densities. In pair plasma, in this particular set-up, the resulting charge separation is negligible.
%Thus, in pair plasma, the release of the magnetic string will mostly generate currents (for \Bf\ along $y$ the currents will be along $z$). Currents $J_z$ will affect $B_y$. It is the dynamics of the $J_z-B_y$ interaction we are after (we work in 1D). 

\subsection{Plasma and  code parameters}

In addition to the 
scale of \Bf\ variation $\lambda$ and the relative amplitude of the \Bf\  fluctuations $ \delta $, one needs to specify the plasma density. We employ the following parameterization (suitable for the laser-oriented PIC code {\it EPOCH})
\ba &&
n_{\rm cr} = \frac{ m_e c^2 }{e^2} \frac{\pi}{\lambda ^2}
\nn &&
n ={\cal N}   n_{\rm cr}
\nn &&
{\cal N} =  \left( \frac{n}{n_{\rm cr}} \right)
\nn &&
\om_{cr} = 2 \pi  \frac{c}{\lambda}
\nn &&
\omega_B = \frac{ e B_0}{ m_e c} \equiv b_0 \om_{cr} 
\ea
In these notations, $b_0=1$ corresponds to the case where the wave with wavelength equal to the $x$-scale of variation $\lambda$ is in cyclotron resonance. In application to magnetar \mss, $b_0 \gg 1$ (in other words, cyclotron radius much smaller than spatial variations of the \Bf).
(We define density (and plasma frequency)  with respect to the density of each component independently, so the total density is $2n$.)

To establish fiducial values, we note that for a wave of amplitude $\delta\approx 1$ the  condition for current starvation is then approximately
\be 
\frac{ \delta B_0}{\lambda} = \frac{4 \pi}{c}  2 n e c \to
 b_0 = 4 \pi {\cal N}   
\label{charge-starv}
\ee
The plasma magnetization parameter is 
\be
\sigma = \frac{\om_B^2}{\om_p^2} = 
\frac{b_0^2}{{\cal N} }
\label{sigma}
\ee
Combining the definition of sigma-parameter (\ref{sigma}) and the condition for charge starvation
(\ref{charge-starv}) gives a special value of sigma
\be 
\sigma ^\ast = 4 \pi b_0 =
16 \pi^2  {\cal N} 
\label{sigmastar}
\ee
It is special in a sense  that a nonlinear wave with $\lambda$ and  $\delta \approx 1$,  propagating in plasma with ${\cal N}$ will be marginally current-starved.

For the basic  runs,  we use $L = 10 \lambda$, $n_p=100$ (particles per cell), $n_x=10^5$ 
(total number of cells, so  $10^4$ cells per disturbance length $\lambda$, and we run for $\approx 10 \times c/\lambda$ dynamical times.
We set ${\cal N} =10^3$ (so that plasma density is much higher than the critical), and $\sigma = \sigma ^\ast$. The resulting \Bf\ parameter is
\be 
b_0 = 4 \pi {\cal N}  = 1.2 \times 10^4
\ee

As we  observe efficient particle acceleration, an important parameter is the maximal expected \Lf\ (same as 
Hillas's criterion)
\be
\gamma^\ast = \frac{e B_0 \lambda}{m_e c^2}= 
2 \pi  b_0 = 7.8 \times 10^{4} \gg 1;
\label{gammaast}
\ee 
the maximal \Lf\ that we observe  in simulations is $\gamma_{max} =  4.2 \times 10^{4}$, consistent with (\ref{gammaast}).

\section{Results of simulations}

\subsection{Overall evolution}

\begin{figure}[h!]
 \includegraphics[width=.99\linewidth]{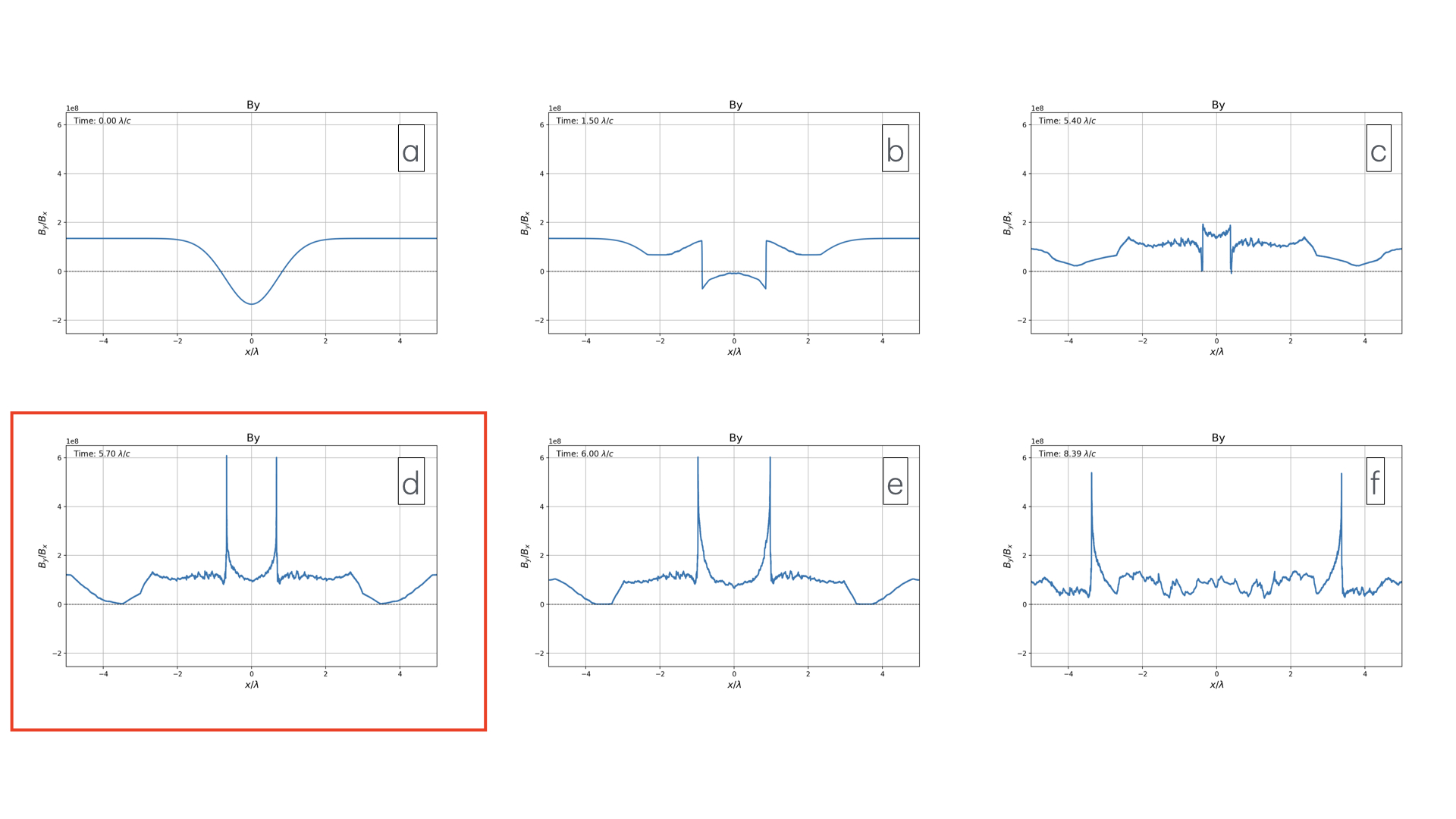}
\caption{Evolution of \Bf\ $B_y$ (same times as Fig. \ref{Extracted_Frames_Panel_log}. In panel c, \Bf\ is zero at points $x\approx \pm  \lambda/2$. In  just $0.3 \lambda/c$, the \Bf\ is strongly enhanced within the collapse region. Parameters: $\sigma = \sigma^\ast$, $\delta =2$. 
} 
\label{Extracted_Frames-By}
\end{figure}

In this section we discuss our key results - formation of a short, high frequency bright pulse during X-mode collapse/overturn - using a number of measured quantities.  In Fig. \ref{Extracted_Frames-By}
(corresponding movies can be found in the Ancillary Folder ./anc)
we plot the profiles of the \Bf\ $B_y(x)$.  Panels c-d-e are time-zoomed to the moment of collapse. The first post-collapse frame d is highlighted in red.

 \begin{figure}[h!]
  \includegraphics[width=.99\linewidth]{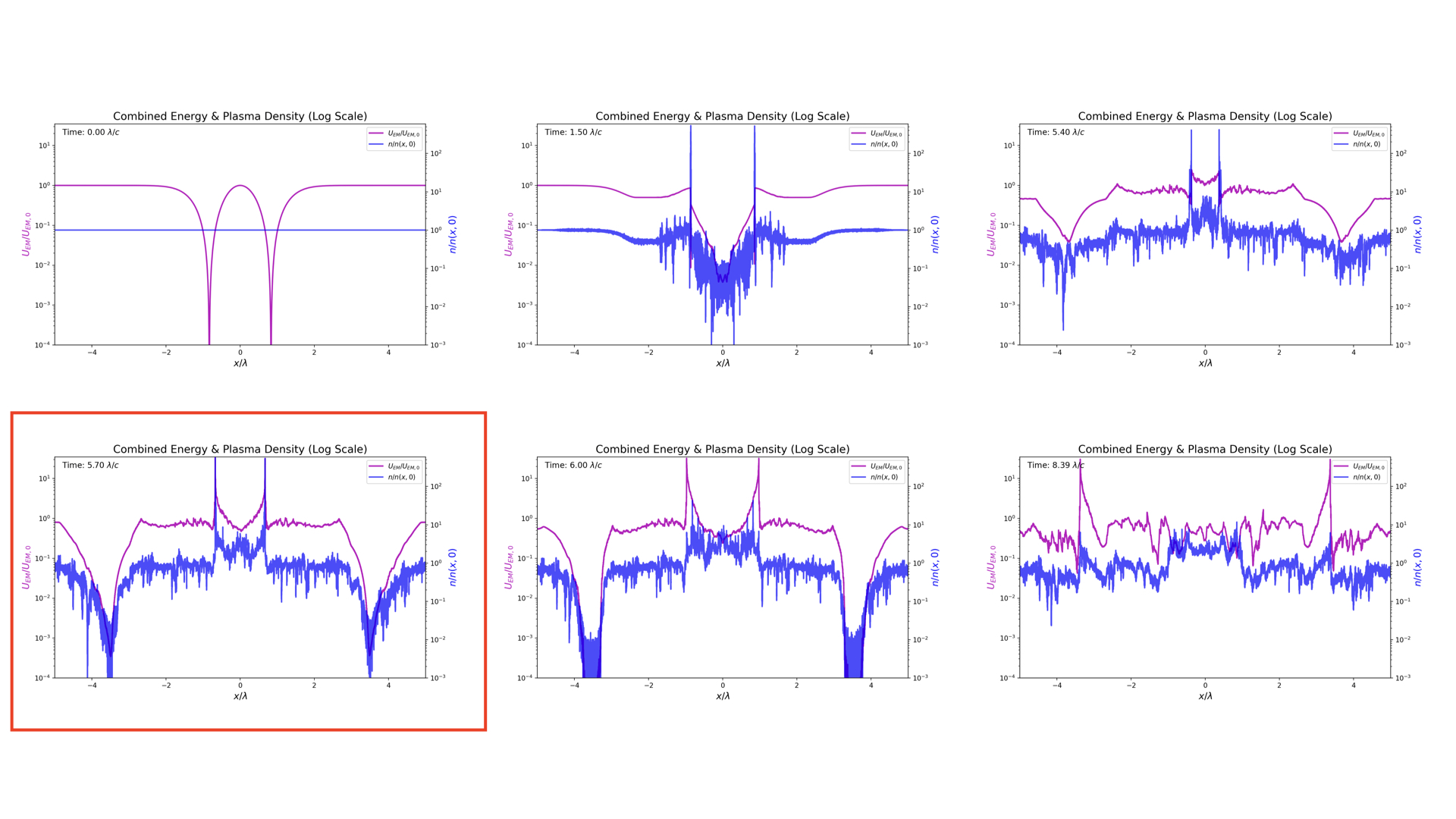}
\caption{Time evolution of density and \EM\ energy density in logarithmic scale; 
}
\label{Extracted_Frames_Panel_log}
\end{figure}

In Fig. \ref{Extracted_Frames_Panel_log} we combine plots of plasma density and \EM\ density (in logarithmic scale). Near-collapse (frames c-d-e) highlight energy depletion from the bulk. Energy dips are {\it co-spatial} with the density dips.

An initial wave separates into two counter-propagating waves. That quickly creates trapped particles, a large density  layer (frame b). Density peaks are  limited only by resolution.
A large density cavity traps the EM inside (frame c). Two counter-streaming EM Poynting fluxes are formed:  from the  inside and another from  outside (frame d). During collapse, particles in  the  large density walls are coherently  accelerated by the EM waves and, in turn, produce a short  pulse of coherent emission (frame d). The pulse propagates as a wave (frame f).

\begin{figure}[h!]
  \includegraphics[width=.99\linewidth]{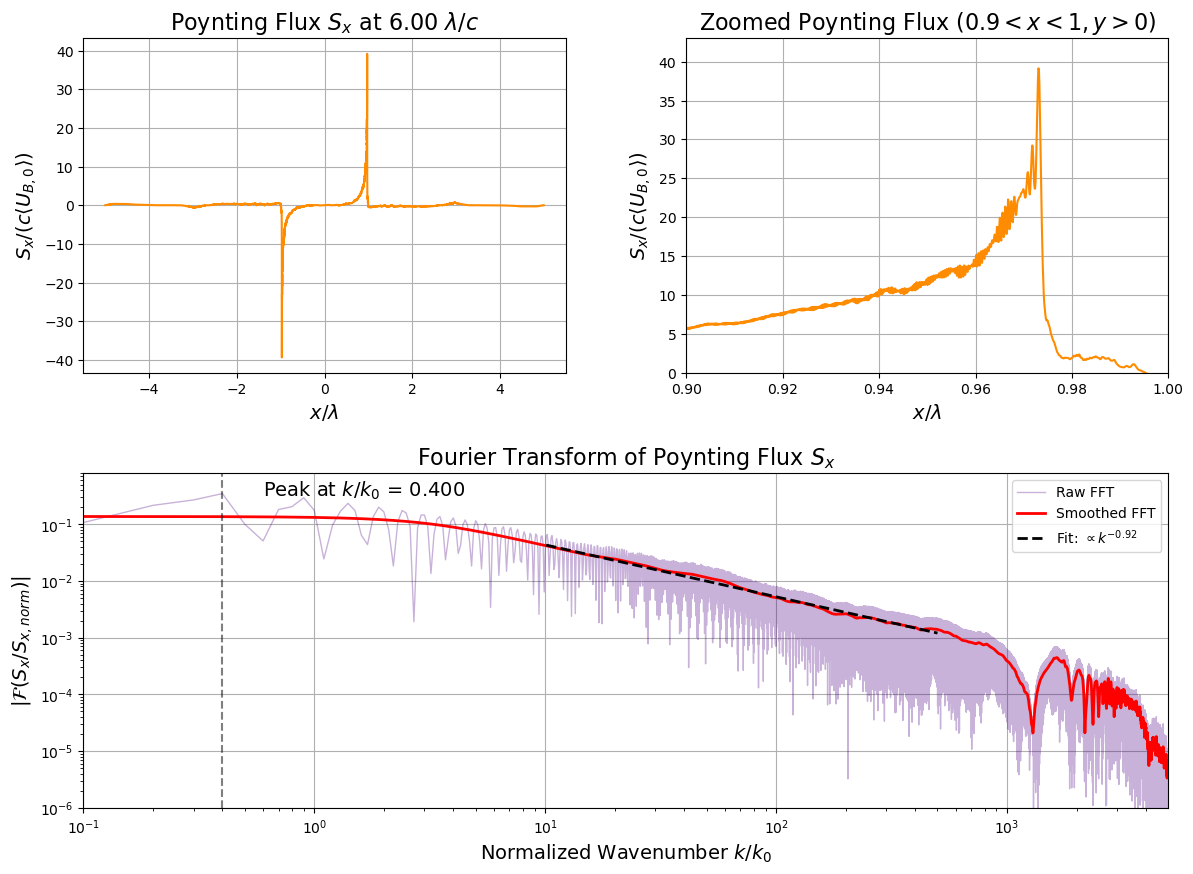}
\caption{
Poynting flux  profile (top row) and Fourier transform (bottom). (Spectral features at $\sim 2 \times 10^3 k_0$ is likely a numerical artifact. Top right panel  is  a zoomed-in view of the peak, showing that we resolve the spike.   } 
\label{Poynting_fft_0020}
\end{figure}

Fig. \ref{Poynting_fft_0020} (top row) shows late-time Poynting flux profile (including the zoomed-in view of the newly generated pulse).  
Fig. \ref{Poynting_fft_0020} bottom row shows the Fourier transform of the Poynting flux. With the resolution of $10^4$ cells per $\lambda$ we observe generation of $k \sim 10^3 k_0$ over just 6 dynamical times $c/\lambda$. 

 The amplitude spectrum  is approximately $\propto k^{-1}$ (bottom panel); hence  the power spectrum is $\propto k^{-2}$. The peaks are well resolved overall, yet the hard spectrum likely  indicates  singularity.

\begin{figure}[h!]
  \includegraphics[width=.49\linewidth]{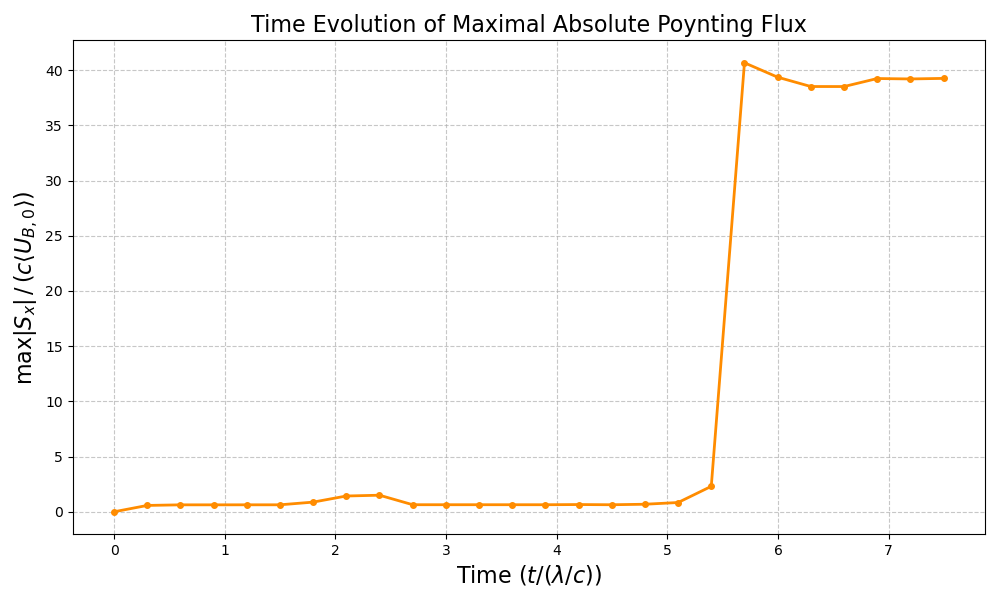}
   \includegraphics[width=.49\linewidth]{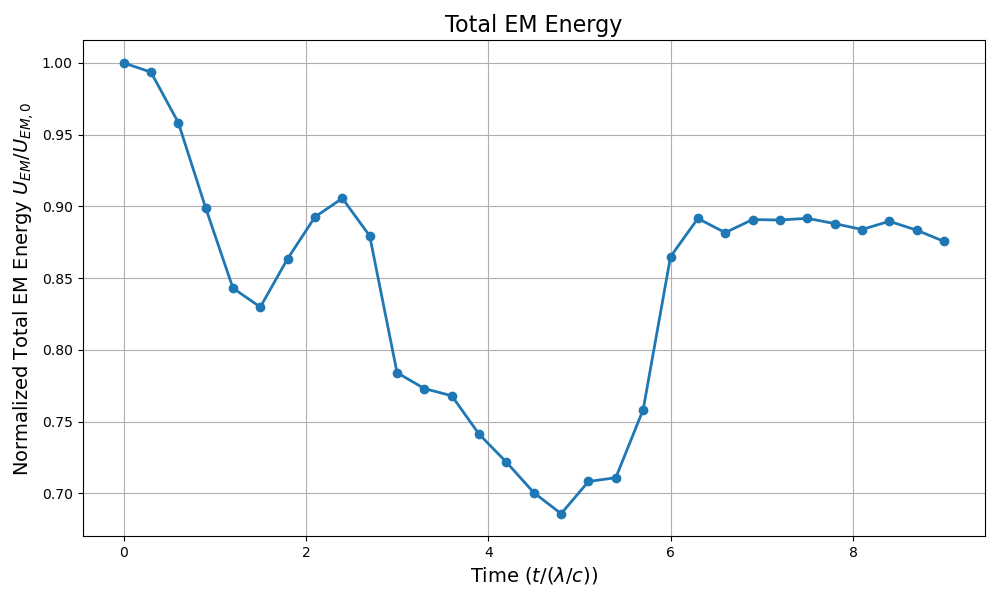}
\caption{ Left: Time evolution of maximal Poynting flux. Formation of a bright pulse is clearly seen. Right: Evolution of the EM energy. Around time $3 c /\lambda$ EM energy is given to the particles of the dense wall, which quickly re-emit it  around time $4 c /\lambda$.
At late times nearly 90\% of the initial \EM\ energy is concentrated in the short pulses.
} 
\label{Max_Poynting_Flux_Evolution}
\end{figure}

Next, in Fig. \ref{Max_Poynting_Flux_Evolution} we plot time evolution of the  maximal Poynting flux and the total \EM\ energy. The formation of a bright pulse is clearly seen. 

Also, after the pulse formation the Poynting flux remains constant, while pulses propagate without much change, keeping the spectra approximately constant. This indicates   that the pulse propagates without dissipation: it's a high frequency X-mode pulse! {\bf This is the FRB.}

The evolution of the \EM\ energy is quite revealing: after mild dissipation  at times $\leq c/\lambda$,  the \EM\ energy  recovers. {\bf At later times,  the short  pulse carries a large fraction of the total  distributed initial \EM\ energy.}

%Fig. \ref{Extracted_Frames_JzEz} illustrates the energy exchange between  plasma and particles. During collapse $J_z E_z \leq 0$, indicating energy transfer from fields to particles. At the same time, energy of the fields increases???

\begin{figure}[h!]
  \includegraphics[width=.9\linewidth]{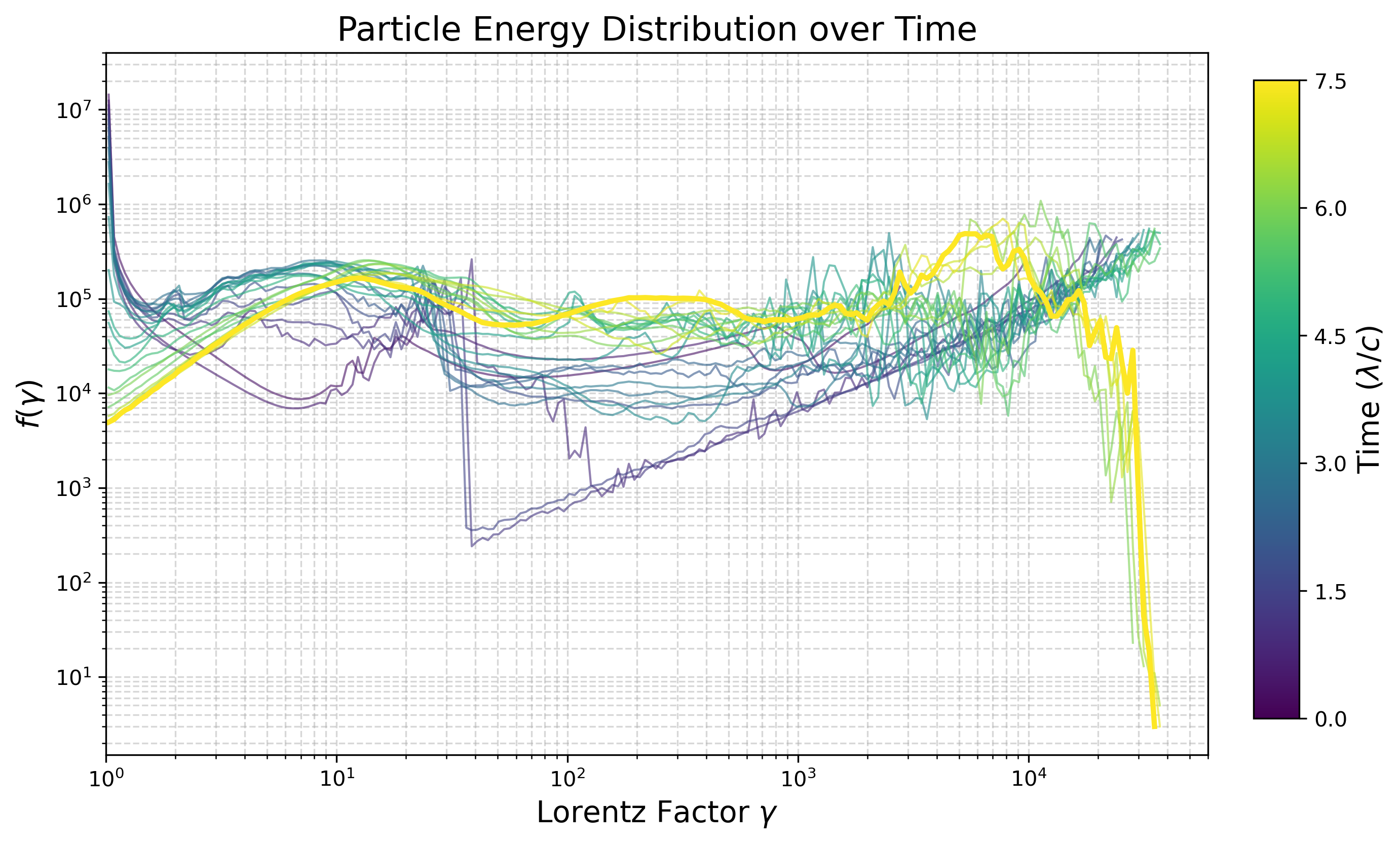}
\caption{Evolution of  distribution function.
} 
\label{fgamma-final}
\end{figure}

In Fig. \ref{fgamma-final} we plot the evolution of the distribution function.  The distribution function $f(\gamma)$ remains mostly flat $f(\gamma) \propto \gamma^0$ from $\gamma \approx 1$ to  $\gamma_{\rm max} \approx 4.2 \times 10^4$. There is an appreciable dip, by two orders of magnitude,  around time $\approx c/\lambda$  at  $\gamma \approx 40$; the origin of this feature is not clear.

\begin{figure}[h!]
  \includegraphics[width=.9\linewidth]{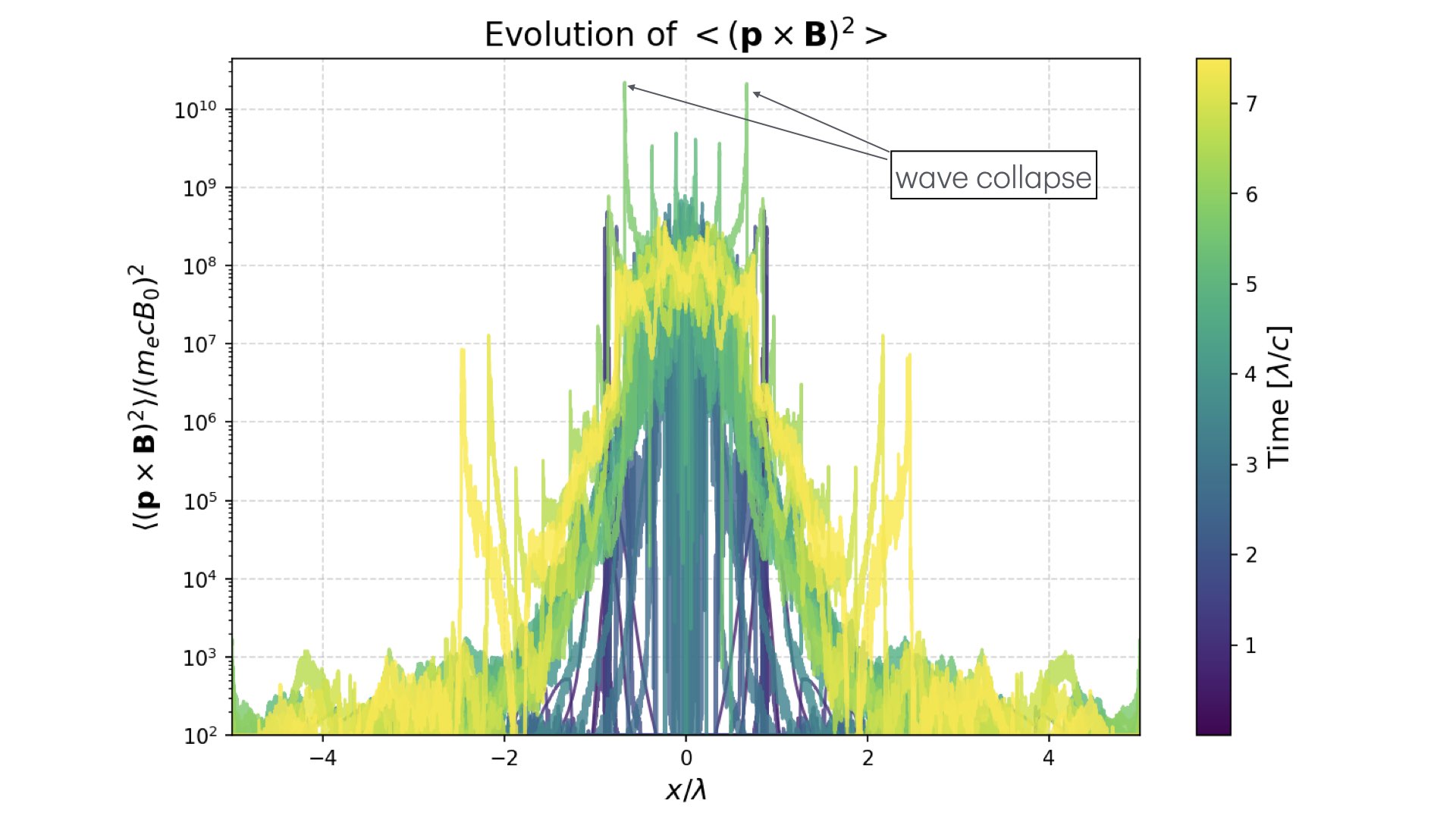}
\caption{Evolution of local averaged values of $< ( {\bf p} \times {\bf B})^2>$, normalized to $(m_e c B_0)^2$. This serves  as a proxy for radiative losses. 
} 
\label{pcrossB}
\end{figure}

To access the importance of radiative loses, in Fig. \ref{pcrossB} we plot averaged values of $< ( {\bf p} \times {\bf B})^2>$ (in 1D set-up, for the  X-mode, the momentum is actually always perpendicular to the magnetic field). One can clearly identify the spikes at the moment and location of wave collapse. 

Though the absolute importance  of radiative losses depends on the
 actual values of fields and particles' momenta, 
Fig. \ref{pcrossB} demonstrates  that overall losses are highly concentrated at the moment/location of the wave collapse. One expects then an associated burst of high energy emission contemporaneous with an FRB.
 
%The plots of $j_z E_z$ (not shown here) show a brief mild positive spike, indicating energy transfer from \EM\ the plasma plasma particles (as expected at the shock).

\subsection{First time steps: trapped particles, phase mixing  and  formation of little monster shock} 

\begin{figure}[h!]
  \includegraphics[width=.99\linewidth]{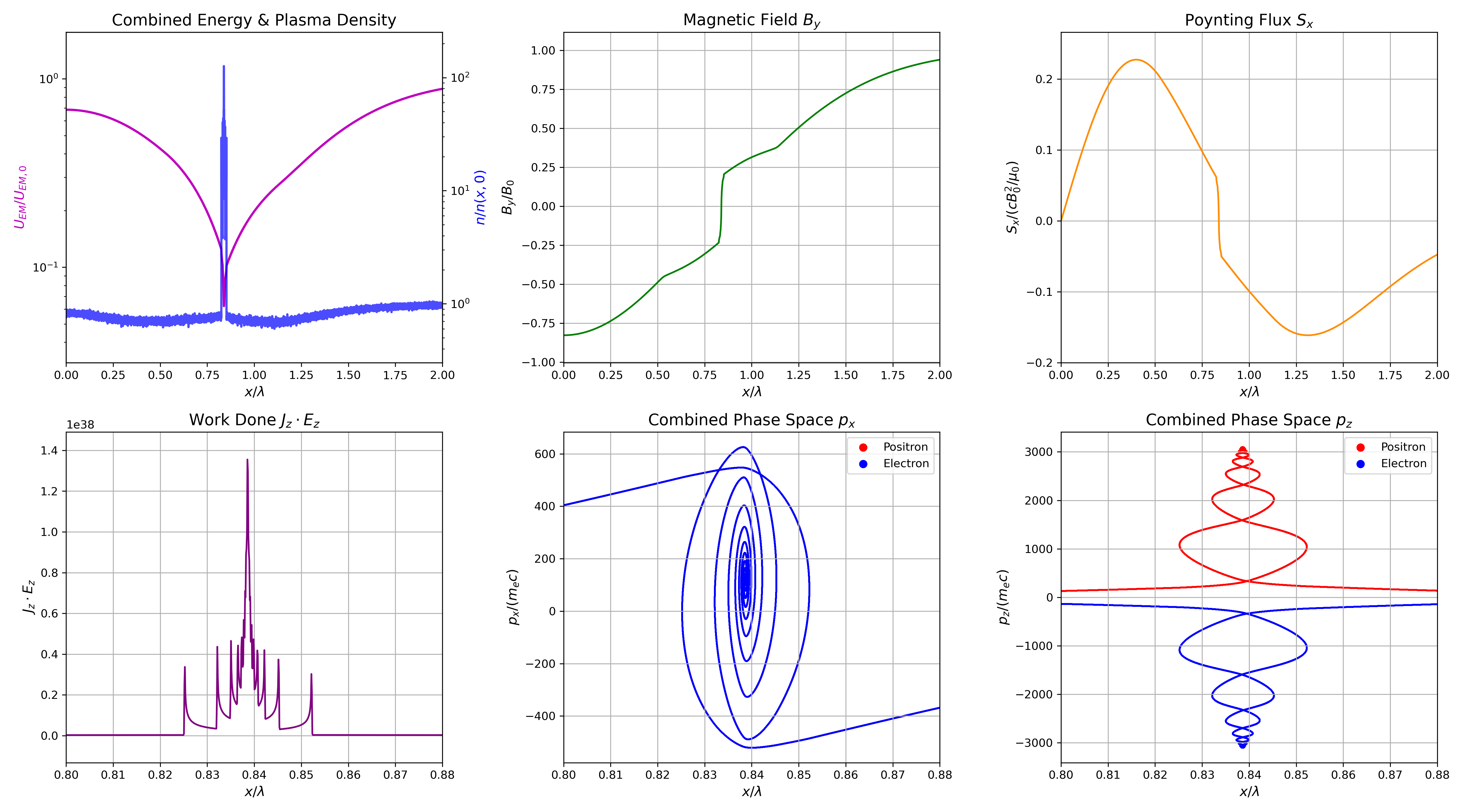}
\caption{First step:  towards formation of a little  ``monster shock''. Top panels: Combined density and energy density, \Bf\ $B_y$ and Poynting flux $S_x$. Bottom panel: zoomed-in view of $J_z E_z$ and particles' phase portraits $x-p_x$ and $x-p_z$ (the tracks on $x-p_x$ plane are identical for two species).
} 
\label{Panels_0001}
\end{figure}

In Fig. \ref{Panels_0001} we plot plasma parameters at out first print step at  $0.3  \lambda/c$.
 From Fig. \ref{Panels_0001} (bottom middle and left) it is clear that particles are not just accelerated to a particular value of momenta: there are intricate phase trajectories. 
Formation of trapped particles is  a common feature of nonlinear wave interaction \citep[eg, Fig. 7  in Ref. ][]{2025arXiv250920594L}. 

%Regarding the formation of  ``monster shocks", it is incorrect to assume that vacuum addition of waves with $|E| \geq |B| $ during some phases of the waves  produces acceleration and wave dissipation. The counter-arguments are many. First, any superluminal wave, e.g. the basic EM wave in plasma has $|E|  > |B|$ - particles are not accelerated.  In plasma,  the current reacts both to electric and magnetic field, in a collective manner: vacuum logic cannot be used.Only in constant electric and magnetic fields  the drift velocity becomes superluminal for $|E| >|B|$. Generally, as the name signifies, drifts are averaging over some fast periodic motion (E-cross-B  drift in {\it constant } fields is an exception). 
 
 We do observe instances of  plasma-EM wave {\it reversible} energy exchanges due to  intricate phase trajectories (not just  E-cross-B drift).
 Eventually,  the phases become mixed -  that's when one can claim a shock forms.
 %, Fig. \ref{phase_movie}.
 Importantly, for  our basic example, only a minor part of the wave's energy is transferred to the  particles.

 %plot_panel_0001.py

% We do track the EM invariant $B^2 - E^2$ - it remains mostly positive.

\subsection{Zooming-in to the collapse moment} 

The X-mode collapse in our simulations occurs near time $5.7 c/\lambda$. 
(Since for our parameters the collapse time is longer than the period, a harmonic wave does not experience powerful breaking.)
Then, in one time step ($0.3 c/\lambda$) extremely powerful {\it high-frequency} \EM\ pulse is generated. At peak value the Poynting flux is $\sim 40 \times$ initial magnetic energy density (times speed of light).  Overall the short  pulse carries $\sim 20\%$ of the initial {\it total} \EM\ energy.

\begin{figure}[h!]
   \includegraphics[width=.99\linewidth]{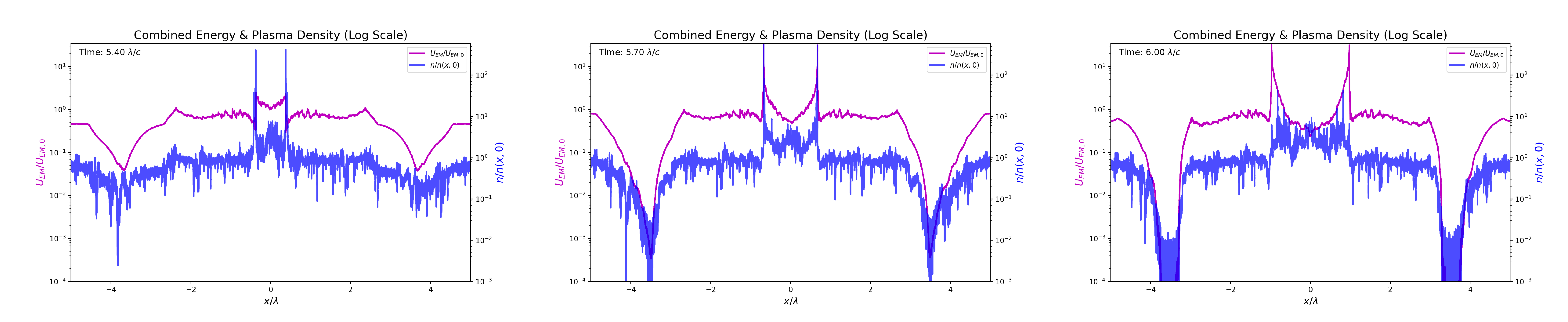}
     \includegraphics[width=.99\linewidth]{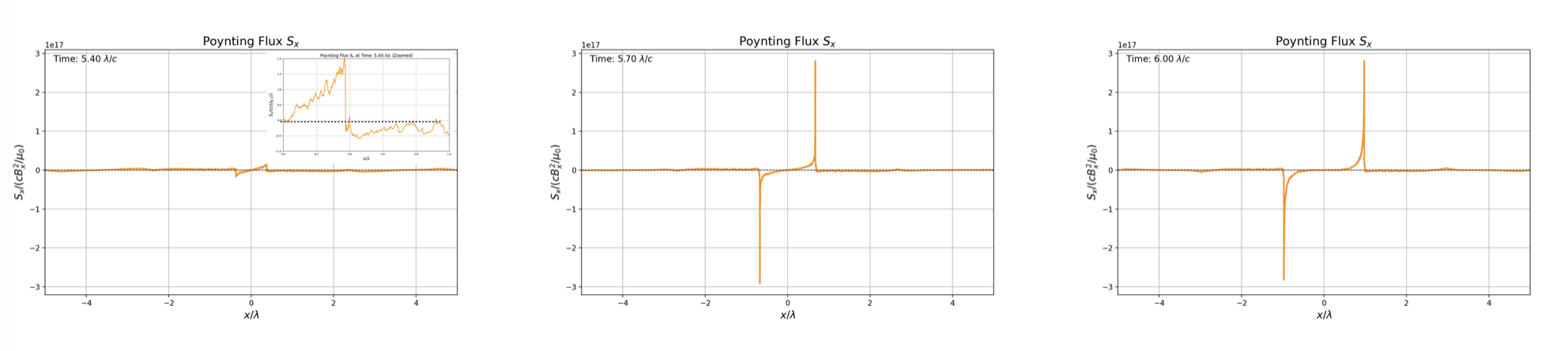}
       \includegraphics[width=.99\linewidth]{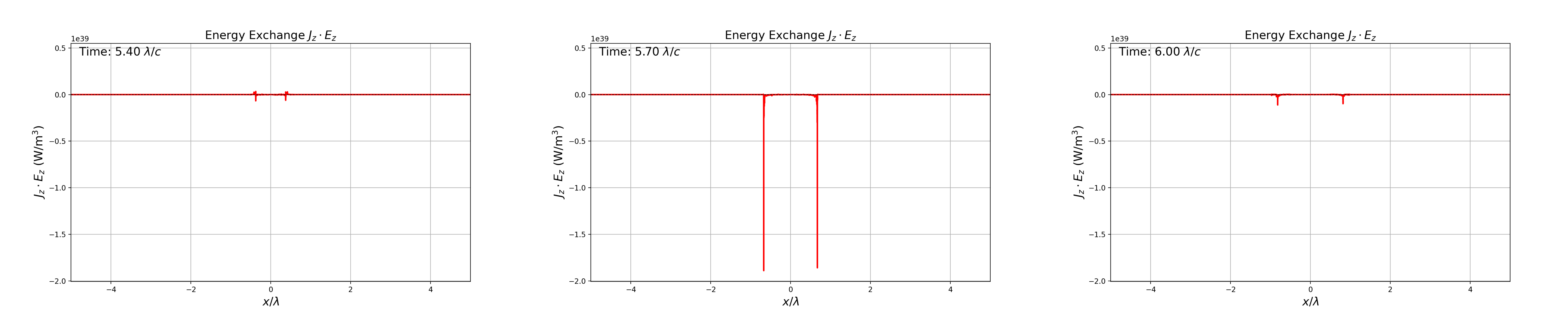}
    \includegraphics[width=.99\linewidth]{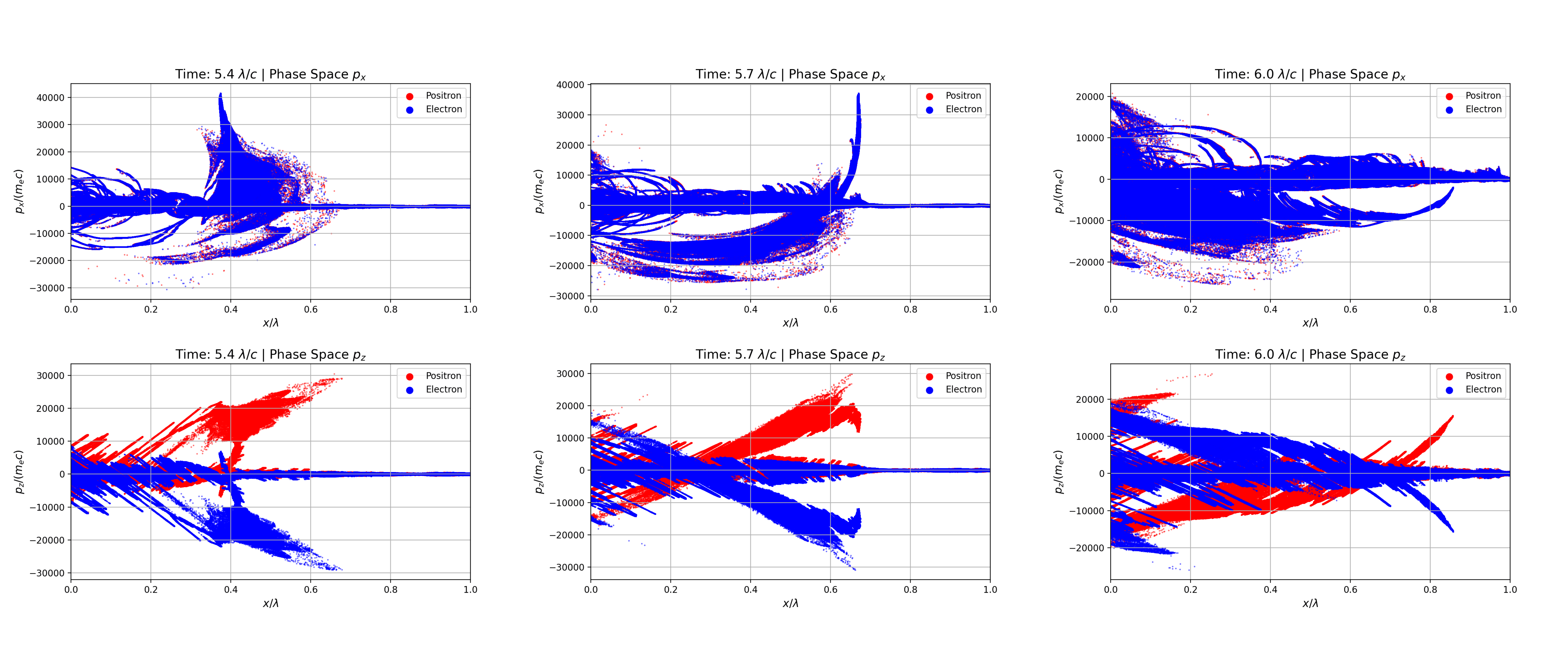}
\caption{ The moment of X-mode collapse. Three consecutive time steps: right before (left Column), during (middle column), and after collapse  (right  column). Plotted are: combined density and \EM\ energy density (top row), Poynting flux (second row; the insert in the left pre-collapse frame demonstrates that before the collapse the energy is flowing  from both sides towards the future collapse location), plasma-particles energy exchange  $J_z E_z$,  and phase space maps (two  bottom rows). Notice sudden cooling of the counter-propagating pair beams, as energy is transferred from particles to waves (bottom row, middle and right panels).
} 
\label{phase_movie-collaps}
\end{figure}

Before collapse, one observes formation of two counter-streaming (along $z$ direction) particle beams, which then quickly lose energy (second row in Fig. \ref{phase_movie-collaps}.)

The massive jump in the local field amplitudes during X-wave collapse is a classic signature of a nonlinear spatial wave collapse (or wave breaking).   The catastrophic spike in the local electromagnetic energy density ($U_{EM}$), peak magnetic field ($B_y$), and Poynting flux ($S_x$) is \emph{not} caused by new energy entering the system or a numerical instability. Integration over the simulation domain reveals that the total global electromagnetic energy remains nearly constant. 

The electromagnetic wave is undergoing severe spatial steepening.  Driven by nonlinear modifications to the plasma refractive index and strong ponderomotive forces, the wave packet rapidly longitudinally  self-focuses, compressing its physical width down to a small scale (yet, still well-resolved  at $\sim 100$  cells).
As the spatial width of the wave collapses, the entirety of its stored energy is squeezed into this highly localized macroscopic singularity. 

In summary, the X-mode wave is violently longitudinally self-localizing, forcing the local field amplitudes to skyrocket purely as a consequence of extreme spatial compression and global energy conservation.

%Investigating the $x-p_z$ phase portrait of the system,Fig. \ref{Extracted_Frames_Zoomed_Profiles_movie_new}, one observes rapid cooling of the high energy particles, \eg, above $|p_z| \geq 10^4$. This is the source of energy driving the production of the high frequency pulse. 

\subsection{Variations of parameters}

\begin{figure}[h!]
 \includegraphics[width=.99\linewidth]{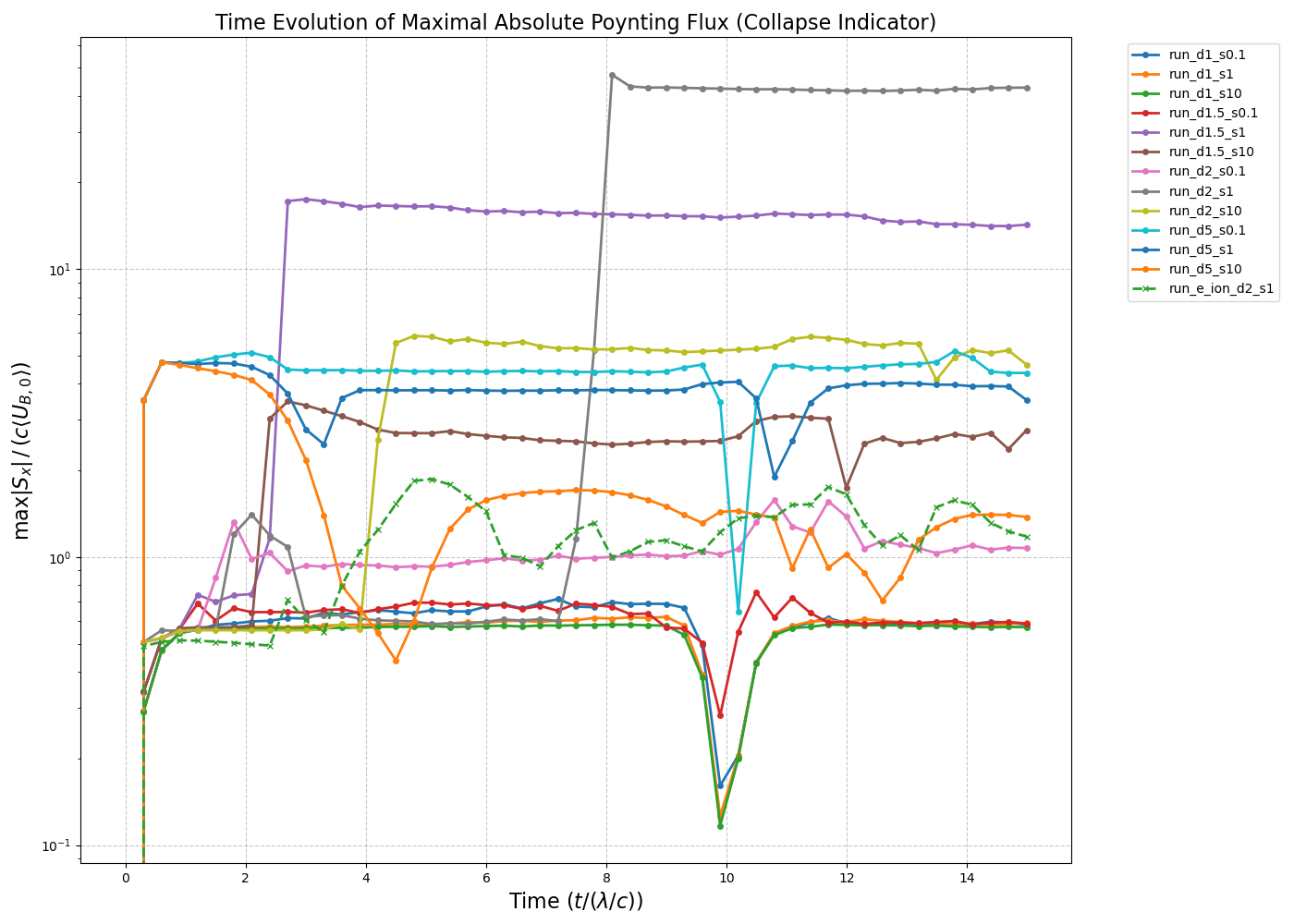}
\caption{Parameter scan: evolution of the maximal value of Poynting flux (logarithmic scale, normalized to the initial \EM\ energy density; curve notations are $d\equiv \delta$, $s_ 0 \equiv \sigma/\sigma^\ast$. A run with fixed ions is also included. Maximal value of maximal  Poynting flux corresponds to $\delta \approx 2$, $\sigma \approx \sigma^\ast$.
} 
\label{sigma05compare}
\end{figure}

The effect of X-mode collapse is highly ``localized'' to a particular parameter space, Fig. \ref{sigma05compare}. A powerful X-mode collapse occurs in a limited region of $\delta \approx 2$ (so that the field varies between $\pm B_0$) and $\sigma/\sigma^\ast \approx 1$ (so the system is near the current-starvation regime).

%sigma05compare

Additionally, we notice:
\begin{itemize}
 \item For higher $\sigma/\sigma^\ast$,  collapse occurs earlier, while the peak Poynting flux, as measured with respect to the initial energy density of the \Bf\ is smaller. 
    \item 
With no reversal, $\delta  \leq 1$ the initial configuration just splits into two counter-propagating waves  
 \item 
Mild temperatures, $\Theta \sim 1$, do not affect the results (as we are in the $\sigma \gg 1$ regime).
 \item  A qualitative changes occurs  $\delta  \geq 2 $ (so that initially the energy density in the center exceeds energy density at the edges). In this case,  two mild    {\it  outgoing} EM waves are launched. This leads to overall equilibration of the system without generation of  powerful pulses.
  \item decreasing resolution to $n_x =10^3$ (hence a 100 times speed-up in computation time) delays the moment of collapse by $\sim 25\%$, leaving the overall properties approximately the same
  \item collapse does not occur in electron-ion plasma
 \end{itemize}

\section{Astrophysical considerations: cosmological Fast Radio Bursts}
\label{astro}

%\subsection{Temporal structure}
Our results have important applications to the physics of Fast Radio Bursts.  Most importantly, we demonstrated that initially macroscopically  distributed \EM\ energy of the nonlinear X-mode is  quickly transferred to small scales/short wavelengths  during the wave collapse.

\begin{figure}[h!]
  \includegraphics[width=.99\linewidth]{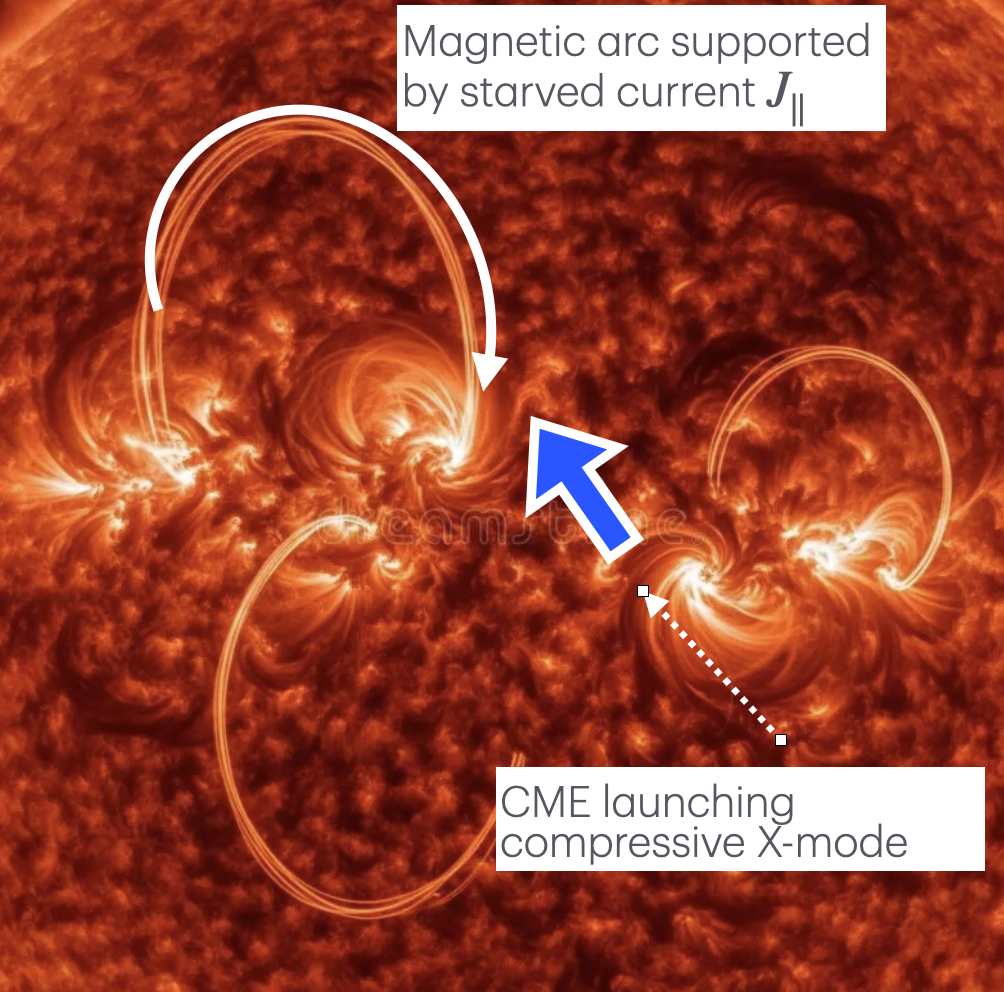}
\caption{Cartoon of possible FRB generation {\it loci}.
} 
\label{cartoonish}
\end{figure}

Fig. \ref{cartoonish}
offers  a cartoonish  description of a possible macroscopic mode. In  a `solar flare' model of magnetar activity \citep{2006APS..APR.X3003L},  a slow evolution of the magnetic field in the upper crust, driven by electron magneto-hydrodynamic flows \citep{1992ApJ...395..250G,2014PhPl...21e2110W,2015MNRAS.453L..93G}, twists the external magnetic flux tubes, producing persistent emission, bursts, and flares \citep{2003MNRAS.346..540L,2015MNRAS.447.1407L}. 
Launching of  Coronal Mass Ejection (CME) generates powerful compressible  X-modes. Especially drastic effects will occur during opening of the \ms\  by the accompanying \Alfven waves \citep{2023MNRAS.524.6024S}.

FRBs of $\sim $ millisecond duration are composed of $N_s \sim 10^3$ sub-bursts of micro-second duration \citep{2023NatAs...7.1486S}. Spatial scales  corresponding to micro-second travel time are of the order of the radius of a \NS; the overall duration then reflects the activity time of the source.

For numerical estimates, 
assume then that an FRB of flux $F_\nu \sim 1 $ Jansky, is split into $N_s \sim 10^3$ sub-bursts, coming from $d=1$ Gpc, and originating near the surface of a magnetar with quantum  poloidal \Bf.  
\citep[Surface \Bf\ are likely to be  much larger, with a contribution from  the toroidal component][]{2006A&A...450.1097B,2009MNRAS.397..763B,2006RPPh...69.2631H}. 

Each sub-burst then carries $\sim 10^{36} $ ergs and requires dissipation of $15$ meters cubed of magnetic energy \citep{2020arXiv200505093L}. Or, an active region of volume $\sim R_{NS}^3$ needs to dissipate $\eta_s \sim 3 \times 10^{-9}$ fraction of energy to generate one sub-burst (and $\eta \sim 3 \times 10^{-6}$ to generate one FRB). The duration of   of each sub-burst matches approximately the dynamical time of a \NS, $\sim R_{NS}/c$. 
(In our simulations, about $20\%$ of the fluctuating part of the  initial magnetic energy is converted in the EM  pulse.)

We have demonstrated that the X-mode collapse produces $k \geq 10^3 k_0$ modes, and even higher. In case of a \NS\ \ms, initial perturbation of $\sim $ kilometers, would produce $\sim$ meters, and shorter wavelengths.

%\subsection{Special configurations}

Importantly, the mechanism operates in a  limited parameter regime, as we have assumed minimal current starvation density, corresponding to  $\sigma ^\ast$, Eq. (\ref{sigmastar}). 
We demonstrate next, this is exactly the regime expected in magnetar \mss.

Magnetospheres of magnetars are twisted \citep{tlk}. For a size of active region $\lambda$ (can be smaller than $R_{NS}$) with mildly twisted \Bf\ $B_0$,  the required plasma density to sustain the twist is given by (\ref{charge-starv}). 
Thus, in the initial configuration, before perturbation,  magnetar plasma is border-line current-starved. 

In addition to supporting the global twist,
the plasma needs to support the current demanded by the perturbations. Qualitatively, any perturbation with amplitude $\delta \sim 1$ will be current-starved, and will experience collapse  producing an FRB.

 In  a `solar flare' model of magnetar activity \citep{2006APS..APR.X3003L},  a slow evolution of the magnetic field in the upper crust, driven by electron magnetohydrodynamic flows \citep{1992ApJ...395..250G,2014PhPl...21e2110W,2015MNRAS.453L..93G}, twists the external magnetic flux tubes, producing persistent emission, bursts, and flares \citep{2003MNRAS.346..540L,2015MNRAS.447.1407L}. 

 Finally, there are numerous ways the radio waves generated deep within the \mss\ can escape absorption, Appendix \ref{escape}.

\section{Discussion}

The \EM\  wave collapse is a unique feature of pair plasmas experiencing a magnetic field reversal at  a specific value of plasma magnetization, so that the nonlinear wave  drives the local current into the current starvation regime. This regime is naturally occurring in magnetars \cite{tlk}.

The present calculations are highly relevant to the problem of generation of astrophysical Fast Radio Bursts. 
{\it  We start with a distributed macroscopic  low frequency \EM\ energy, which  is then quickly converted into short high frequency pulse of radiation.}

One of the most important results of the present work is: formation of a short, high frequency   EM spike involves $\sim 20\%$ of the initial {\bf total} EM energy (see Fig. \ref{Max_Poynting_Flux_Evolution}, right panel: 20\% is the increase of the \EM\ energy after pulse formation). The original X-mode is a long-wavelength mode, but it collapses to produce short/high frequency pulse, see Fig. \ref{Poynting_fft_0020}.

% Amplitude spectrum  of the Poynting flux  $\propto k^{-1}$ is a clear signature of wave breaking (profile is a step function). 
 %\cite[Not to be confused with Phillips spectrum of deep water surface wave, which break via a spike in profile]

% The present model does not appeal to shock formation (little monster shock is a small sideways story). Shocks are typically highly inefficient in generation of coherent emission, as initial source of energy is first transferred to the bulk motion, then dissipated, while a (hopefully)  small unstable part  of particle distribution is used to generate coherent emission -  every step in such prescription is bound to be of low efficiency). 

 The model stresses kinetic plasma response, as opposed to hydrodynamic one:  large-scale, long-lived $x-p_x-p_z$  phase space correlations are the key to generation of coherent radio emission.

 We  observe a  very hard spectrum of accelerated particles, that carries $\sim 10\%$ of the initial distributed \EM\ energy. These particles  might produce short, millisecond-long,  high energy pulses.

% The model appeals to pure, linearly polarized X-mode propagating across \Bf. Our experience with modeling parallel-propagating modes, both superluminal and \Alfven \citep{2025arXiv250920594L,2025arXiv250917245L,2026arXiv260506085L}, indicates that, qualitatively, the plasma dynamics in those cases is ``slow'',  since currents are generally  compensated by the non-radiative  curl-B term. In the case of X-mode breaking,  large small scale \Ef\ are generated. 

We would like to thank Pablo Bilbao, Jason Hessels, Anatoly Spitkovsky, and participants in the 6th Purdue Workshop on Relativistic Plasma Astrophysics for discussions. 

 This research was supported by NASA grant 80NSSC25K0516 and  in part by grant NSF PHY-2309135 to the Kavli Institute for Theoretical Physics (KITP).

\bibliography{BibTex} 

 \appendix

\section{Escape of generated radiation from the \mss: electromagnetic broom}
\label{escape}

Concerns have been raised that in the framework of  magnetospheric models of FRBs \citep{2002ApJ...580L..65L,2010vaoa.conf..129P,2013arXiv1307.4924P,2016MNRAS.462..941L}, the high power nonlinear \EM\ may not escape, suffering from nonlinear absorption \citep{2023ApJ...959...34B}. As discussed by \citet{2024MNRAS.529.2180L} there are numerous ways for the \EM\ to escape absorption,  due to the plasma streaming away, modification of the \mss\ during flare eruption, as well as ponderomotive effects.

\begin{figure}[h!]
\includegraphics[width=0.5\linewidth]{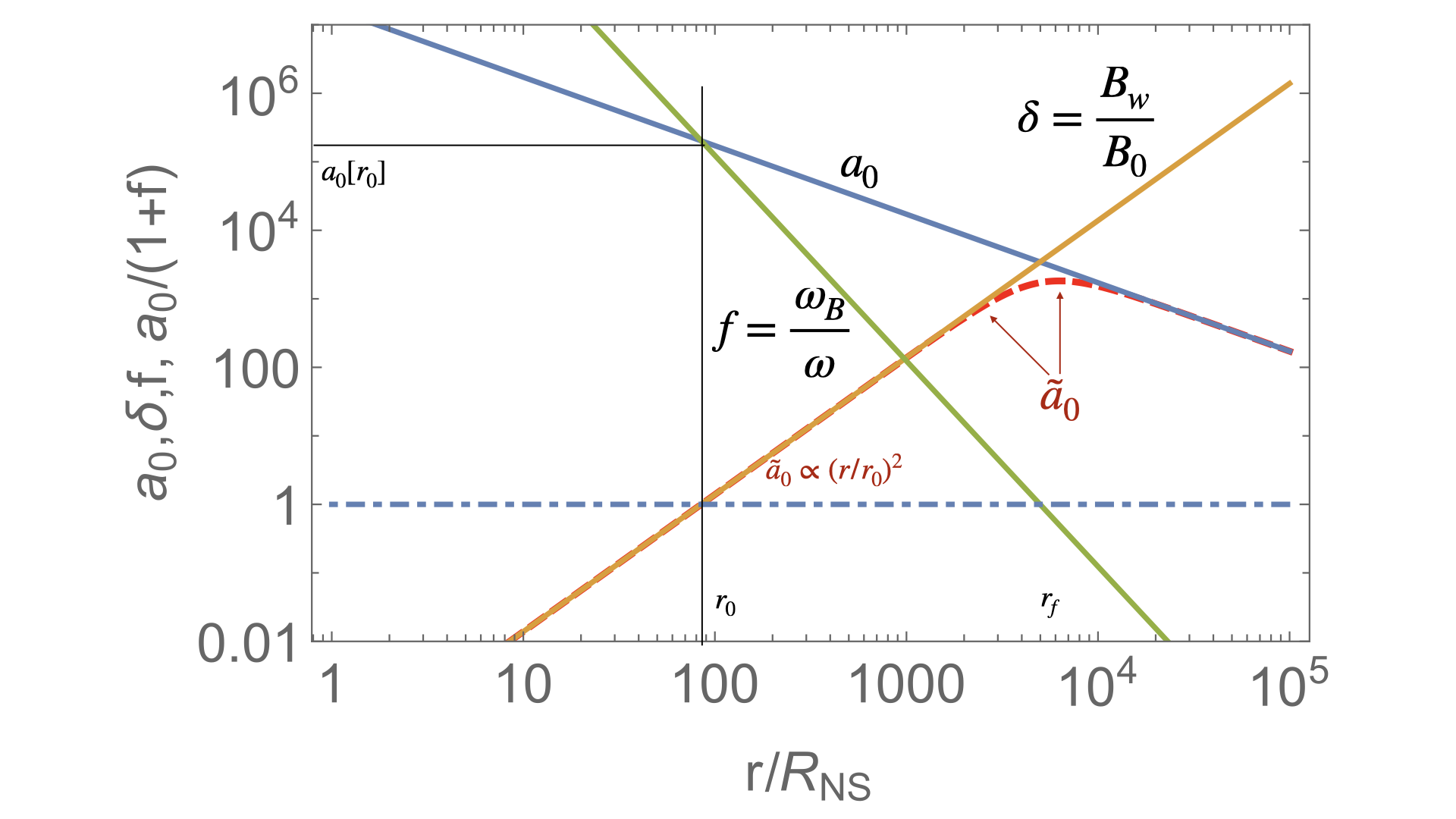}
\vline
\includegraphics[width=0.45\linewidth]{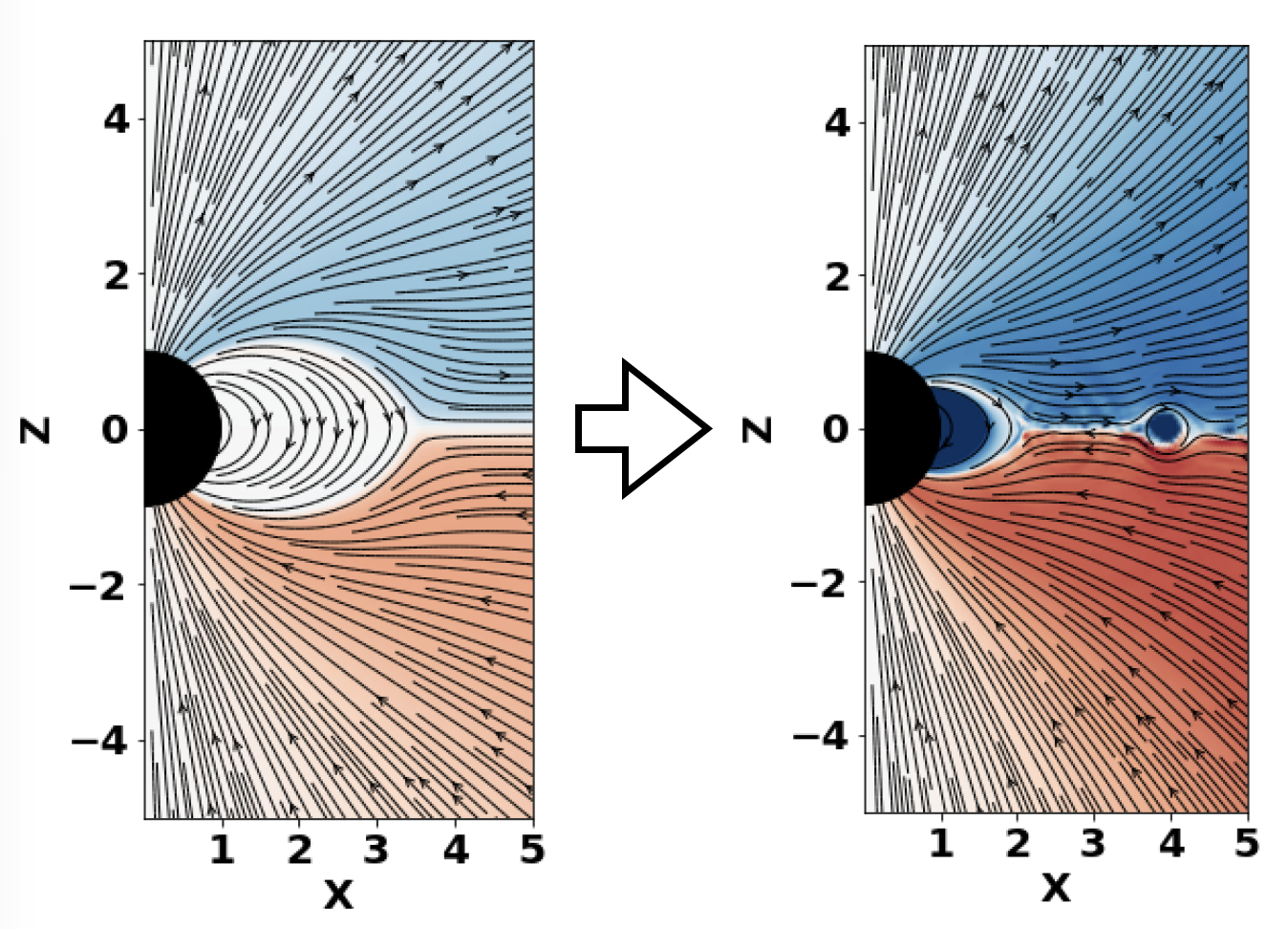}
\caption{ (Left panel) 
 Evolution of basic parameters in the dipolar \ms: nonlinearity parameter $a_0$, relative wave intensity $\delta = B_w/B_0$, ratio of frequencies $f= \om_B/\om$,  and effective   nonlinearity parameter $\tilde{a}_0 = a_0/(1+f)$. Indicated are radii $r_0$ (where  $\tilde{a}_0=1$), value of $a_0[r_0]$, and radius $r_f$ where $f=1$.  Maximal value of  $\tilde{a}_0 ^{(max)} \approx 1.8 \times 10^3 $ is reached approximately at $r_f$.
 Right panel: opening of the \ms\ by the coronal mass ejection: beyond soma radius (also typically $\sim 100$ stellar radii) \Bf\ becomes radial; color is the value of the toroidal \Bf.}
\label{FRB-escapeLC}
\end {figure}

The parameters in magnetar \mss,  Fig. \ref{FRB-escapeLC} left panel \citep[reproduced from][]{2024MNRAS.529.2180L}, demonstrates that in the worst case scenario the  ratio of amplitudes of the wave to guide field starts to become larger than unity only for periods longer than $\sim 30$ milliseconds (recall, Crab pulsar has  a period of $34$ milliseconds). 
Additionally, generation of a CME leads to the opening of the \ms, making \Bf\ lines radial beyond some limit \citep[right panel; adopted from][]{2023MNRAS.524.6024S}.

%FRB-Escape-nonlinear

Another refutable claim is the  absorption in the inner wind, \eg Figure 10 in  \cite{2026ApJ..1000..157B}. The claim relies on the assumption that the wind starts nearly stationary at the light cylinder (Eq. (79)). In fact, the initial relativistic  bulk streaming  of plasma along the field lines  with $\gamma_\parallel \gg 1 $,   makes plasma  to recede nearly  radially  with $\gamma \sim \gamma_\parallel$ (up to $r/R_{LC} \sim \gamma_\parallel \gg 1$
\citep{2022ApJ...933L...6L,2024ApJ...962...18L}. As a result, plasma-wave interaction  will be greatly suppressed by a large power of $\gamma_\parallel$.

\begin{figure}[h!]
\includegraphics[width=0.37\linewidth]{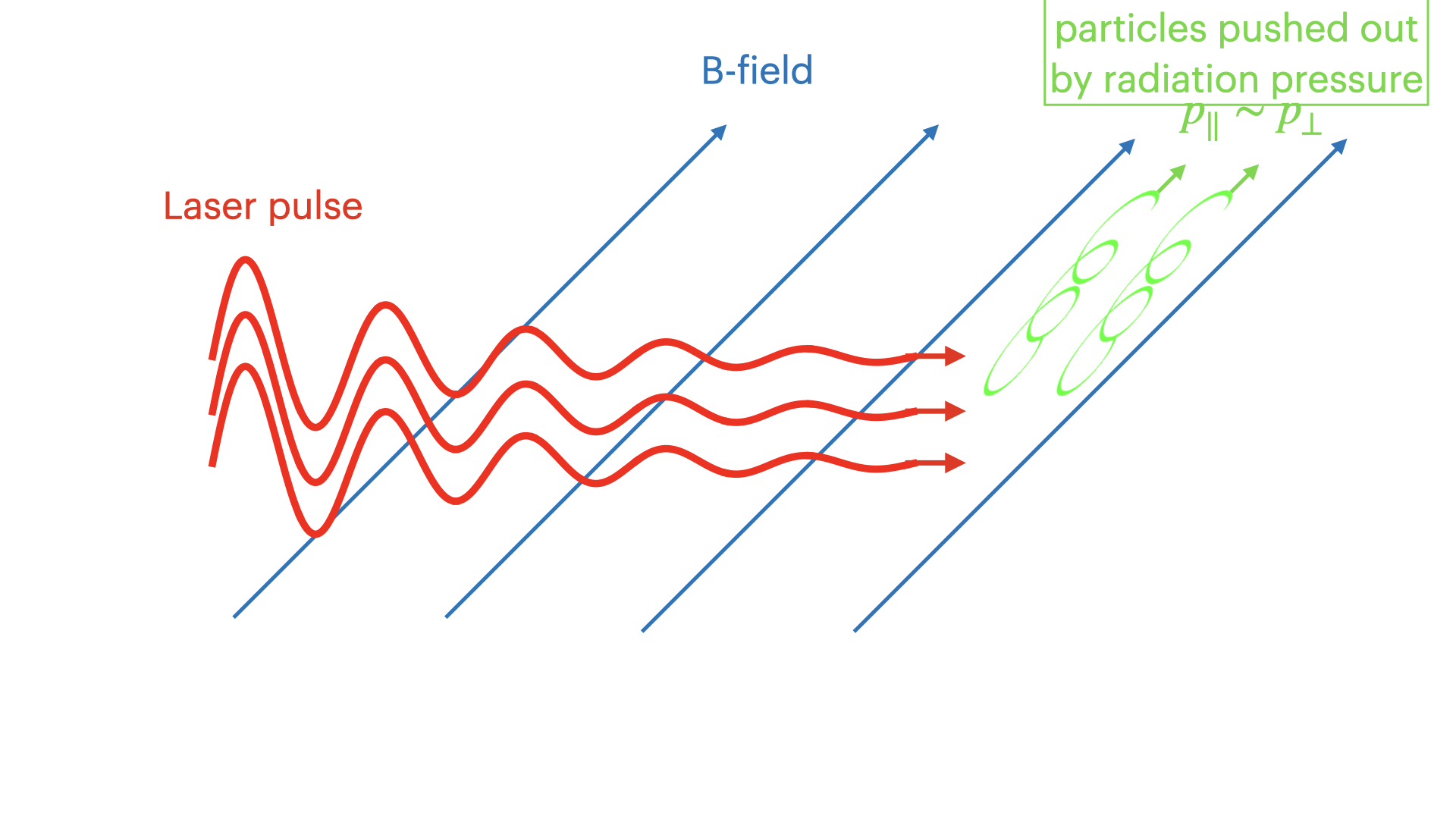}
\includegraphics[width=0.6\linewidth]{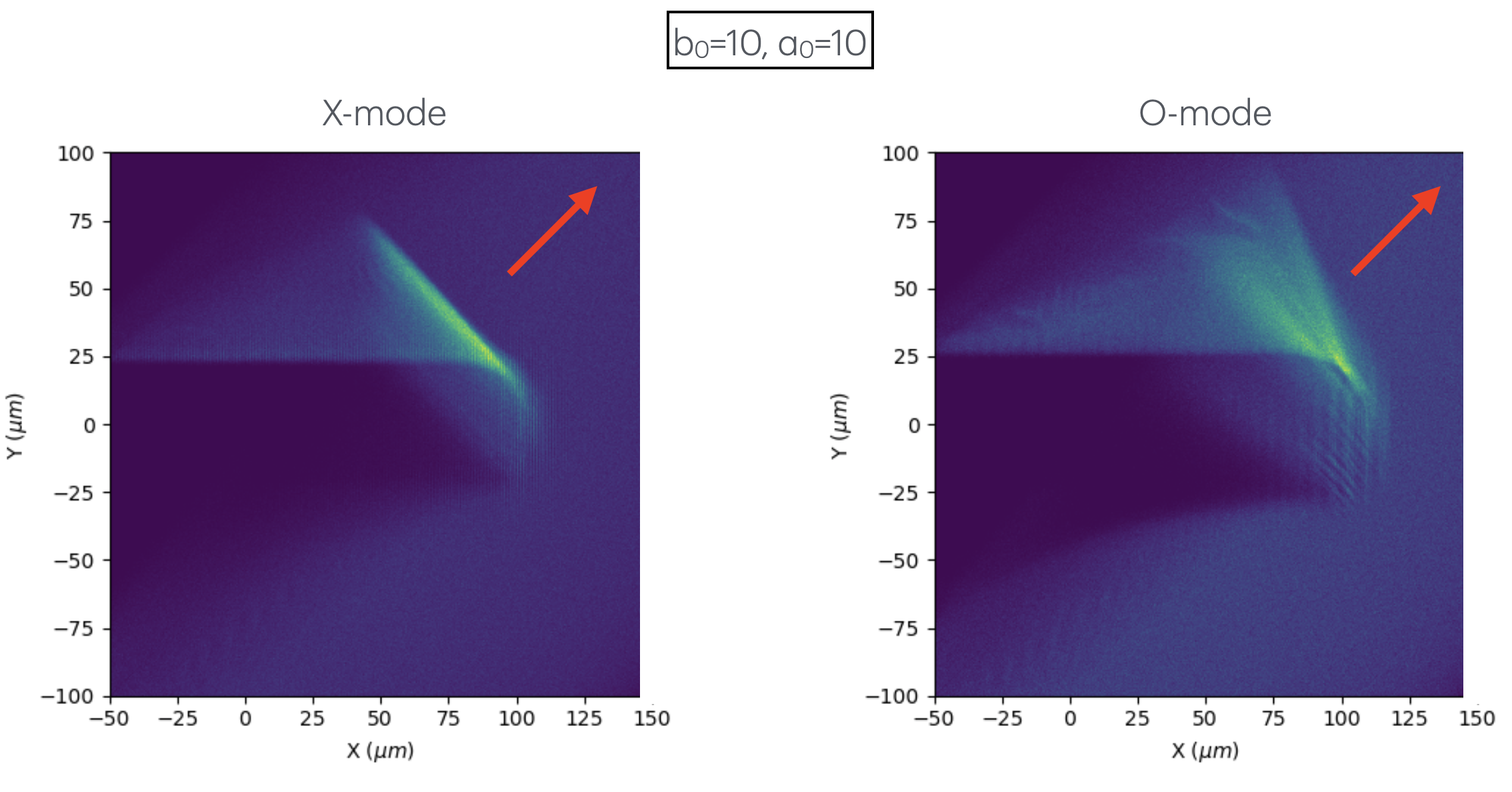}\\
\caption{Electromagnetic broom.  Left Panel: cartoon of plasma's self-cleaning due to the propagation of an intense laser pulse; center and right panel: density maps from 2D simulations of y-modulated linearly polarized Gaussian pulse propagating into pair plasma with oblique \Bf\ (at $45^o$). Laser nonlinearity parameter $a_0=10$, value of the guide field corresponds to the value of the fluctuating \Bf, (parameter $\delta  =1$). The guide \Bf\ is in the plane of the boards as indicated by thick red arrows, X-mode corresponds to \Ef\ of the wave out of the board. The leading part of the pulse sweeps away the plasma, so that the bulk of the EM pulse propagates nearly in vacuum, avoiding possible non-linear absorption. }
\label{2D}
\end {figure}

We  also performed simulations of  a relativistically nonlinear \EM\ wave falling on to plasma with a strong  {\it oblique} guide field,  Fig. \ref{2D}. We observe the \EM\ broom:
 the leading part of the pulse ponderomotively   accelerates plasma particles to $v\sim c$ (but $\gamma \sim 1$) along the \Bf. Thus, the  leading part of the pulse  ponderomotively  pushes the plasma particles along the local \Bf, particles  stream sideways, clearing the path for the main part of the pulse to propagate nearly in vacuum.

\end{document}